\documentclass[epj]{svjour}
\usepackage{epsfig,bm,amssymb,multicol}
\twocolumn
\newcommand{\be}{\begin{equation}}
\newcommand{\ee}{\end{equation}}
\newcommand{\bea}{\begin{eqnarray}}
\newcommand{\eea}{\end{eqnarray}}
\newcommand{\ds}{\displaystyle}
\newcommand{\vep}{{\bm p}}

\newcommand{\vek}{{\bm k}}
\newcommand{\veq}{{\bm q}}

\DeclareMathSymbol{\varGamma}{\mathord}{letters}{"00}

\begin{document}

\title{Interplay of quark and meson degrees of freedom in a near-threshold resonance: multi-channel case}
\author{C. Hanhart\inst{1}, Yu. S. Kalashnikova\inst{2}, A. V. Nefediev\inst{2}}
\institute{Forschungszentrum J\"ulich, Institute for Advanced Simulation, Institut f\"ur Kernphysik (Theorie) and J\"ulich Center for Hadron
Physics, D-52425 J\"ulich, Germany
\and
Institute for Theoretical and Experimental Physics,
117218, B. Cheremushkinskaya 25, Moscow, Russia}

\date{}

\abstract{We investigate the interplay of quark and meson degrees of freedom
in a physical state representing a near-threshold resonance for the case of
multiple continuum channels. The aim is to
demonstrate the full complexity of possible near-threshold phenomena. It turns out
that those are especially rich, if both quark and meson dynamics generate
simultaneously weakly coupled near-threshold poles in the S-matrix. We study
the properties of this scenario in detail, such as $t$-matrix and production
amplitude zeros, as well as various effects of the continuum channels
interplay.
}
\authorrunning{C. Hanhart {\em et al.}}  \titlerunning{Interplay of quark and
meson degrees of freedom in a near-threshold resonance: multi-channel case}
\maketitle

\section{Introduction and Main result}

A convenient parametrisation of an $S$-wave near-threshold amplitude was
suggested in Ref.~\cite{Flatte}. Consider an unstable particle coupled to the hadronic channel, open at $E=0$, with the coupling constant $g_f$. The
$t$-matrix in the effective range (Flatt{\'e}) approximation can be written in the
form:
\be
t(k,k,E)=\frac{1}{8\pi^2\mu}\frac{g_f}{E-E_f+\frac{i}{2}g_fk}\equiv
\frac{1}{8\pi^2\mu}\frac{g_f}{{\cal D}(E)},
\label{FE1}
\ee
where
$$
k(E)=\sqrt{2\mu E}\Theta(E)+i\sqrt{-2\mu E}\Theta(-E),
$$
and $\mu$ is the reduced mass in the hadronic channel. Normalisation of the
$t$-matrix is such that the scattering amplitude is given by
$$
F(E)=-4\pi^2\mu t(k,k,E).
$$
In fact, the Flatt{\'e} form of the scattering amplitude is nothing but the Breit--Wig\-ner amplitude with the momentum dependence
of the elastic width taken into account explicitly.
However, as argued in Ref.~\cite{interplay}, direct interactions between
mesonic channels  could
lead to quite a peculiar form of the line shapes, drastically different from
the ones driven by the simple Flatt{\'e} form (\ref{FE1}).

The simple form (\ref{FE1}) is valid in the near-threshold region, that is, for
\be
|E|\ll|\Delta|,
\label{ED}
\ee
where $\Delta$ denotes the distance to the next threshold,
$$
\Delta=M_{th_2}-M_{th_1},
$$
with $M_{th_i}$ being the $i$-th threshold. However, in practise,
analysis of experimental data requires knowledge of the amplitude in a wider
range, covering a few thresholds. In this case the simple form (\ref{FE1}) has
to be modified. Consider two isospin-related channels. A naive generalisation
of Eq.~(\ref{FE1}) to this case would read\footnote{Notice that, assuming
isospin conservation, we set $g_f^{(1)}=g_f^{(2)}=g_f$ and that here and in
what follows the constant $g_f$ differs by a factor of two from the one
introduced in Eq.~(\ref{FE1}).}:
\be
{\cal D}(E)=E-E_f+\frac{i}{4}g_f(k_1+k_2),
\label{Dsimple}
\ee
where $k_1$ and $k_2$ stand for the momenta related to the corresponding
thresholds:
\bea
k_1(E)&=&\sqrt{2\mu E}\Theta(E)+i\sqrt{-2\mu E}\Theta(-E),\nonumber\\
k_2(E)&=&\sqrt{2\mu E-\Delta)}\Theta(E-\Delta)\label{ks0}\\
&+&i\sqrt{2\mu(\Delta-E)}\Theta(\Delta-E),\nonumber
\eea
and further generalisation to the multi-channel case is trivial. However,
if direct interactions between mesonic channels are present in the system, the
simple form (\ref{Dsimple}) may become inappropriate and thus it may be
subject to generalisation, quite analogous to the one-channel case already
discussed in Ref.~\cite{interplay}. Usually the analysis of experimental data based on the
multi-chan\-nel Flatt{\'e} parametrisation of the near-threshold observables
still employs simple expressions similar to Eq.~(\ref{Dsimple}), with all
effects of the direct interaction between mesonic channels neglected or, more
generally, parametrised into the effective coupling constants --- see,
for example, Refs.~\cite{ourX,ourX2,braatenfw}. It is argued in
Ref.~\cite{braaten2}, however, that the explicit inclusion of such effects
could lead to  line shapes which are even more unusual than those obtained in the one-channel
case discussed in Ref.~\cite{interplay}. The aim of the present paper is to
study systematically the role of the direct interaction between mesonic
channels. In particular, we investigate under which conditions the effect of
direct interactions on the line shape cannot be parametrised into the couplings and has to be taken
into account explicitly. Throughout this paper we assume the constituents of
continuum channels as stable particles. Although a finite width of a
constituent can have a significant impact on line shapes, as already discussed, for example, in  Refs.~\cite{ourX2,braatenfw,finwidth}, this effect can be
treated independently
from the interplay of compact states and continuum channels, which is the
focus here.

As the main quilitative result of our research we find that in most cases one should expect the
multi-channel Flatt{\'e} parametri\-sa\-ti\-on of Eq.~(\ref{Dsimple}) to
describe properly the line shape of near-threshold resonances, however,
under certain conditions (especially for fine tuning of parameters) more
complicated line shapes can emerge. We thus propose the following procedure for the data analysis: 
(i) if the form of Eq.~(\ref{Dsimple}) provides enough structure
to describe new-threshold data, more complicated parametrisations cannot be used in a model-independent way --- the near-threshold data are simply not sensitive enough to discriminate between different parameter sets which provide identical near-threshold behaviour and produce different structures only far beyond the near-threshold region; (ii) if, on the contrary, line shapes in the near-threshold region demonstrate irregularities, more complicated structures, as given
below, should be used. In this case more than two poles are located near the relevant threshold and a non-trivial interplay of various effects is observed. If this would ever be relevant in practice will depend
strongly on how well the background is understood for the particular
measurement; (iii) finally, the mulit-channel formalism developed below may be used to study
the resolutions/statistics necessary for the data to actually detect the
possible interplay of compact states and continuum channels.

The paper is organised as follows. First, in Sec.~\ref{Mcp} we derive general expressions for the
$t$-matrix and the wave function (w.f.) of a near-threshold resonance which appear as a result of
the interplay of a bare quark state and multiple continuum
states. This section is quite technical. It forms the foundation of our
work but can be skipped by those who are more interested in examples of phenomenological applications of the formalism developed. Those are given for the two-channel, near-threshold case in Sec.~\ref{twochan}, which is
then studied in detail from the point of view of the effect of the direct
interaction between mesonic channels. The formulae derived are exemplified
further by the case of the $X(3872)$ charmonium in Sec.~\ref{example}. In particular, we argue that
description of the experimental data for the $X(3872)$ charmonium in the
near-threshold region does not require severe modifications as compared to the
naive approach and can be safely performed with the help of the simple formula
(\ref{Dsimple}). Finally, in the concluding part of the paper, we discuss
various practical aspects of the use of the formulae derived in the data
analysis.

\section{The multi-channel problem}\label{Mcp}

\subsection{Essentials of the formalism}

We study a physical state (for the sake of convenience referred to as $X$)
which is a mixture of a bare $q \bar q$ state (labelled below as 0) and
multiple two-body components (labelled as $i=1,2,\ldots$) and represent its
w.f. as:
\be
|X\rangle=\left(
\begin{array}{c}
c|\psi_0\rangle\\
\chi_1(\vep)|M_{11}M_{12}\rangle\\
\chi_2(\vep)|M_{21}M_{22}\rangle\\
\cdots
\end{array}
\right),
\label{state}
\ee
with $\vep$ being the relative momentum in the corresponding meson pair $\{M_{i1}M_{i2}\}$.

We allow for direct interactions between mesonic channels $i$ and $j$ encoded in the potentials $V_{ij}(\vep,\vep')$ (including diagonal terms with $i=j$).
Transition between the $i$-th mesonic channel and the quark state is caused by the transition potential $f_i(\vep)$.
Then w.f. (\ref{state}) obeys a Schr{\"o}dinger-like equation:
\be
\hat{\cal H}|X \rangle=M|X \rangle,
\label{Sheq}
\ee
with the Hamiltonian
\be
\hat{\cal H}=
\left(
\begin{array}{cccc}
\hat{H}_0&\hat{f}_1[\vep]&\hat{f}_2[\vep]&\cdots\\
\hat{f}_1[\vep]&\hat{H}_{h_1}[\vep,\vep']&\hat{V}_{12}[\vep,\vep']&\cdots\\
\hat{f}_2[\vep]&\hat{V}_{21}[\vep,\vep']&\hat{H}_{h_2}[\vep,\vep']&\cdots\\
\cdots&\cdots&\cdots&\cdots
\end{array}
\right).
\label{H}
\ee

Here
$$
\hat{H}_0|\psi_0\rangle=M_0|\psi_0\rangle
$$
and
$$
\hat{H}_{h_i}[\vep,\vep']=\left(m_{i1}+m_{i2}+\frac{p^2}{2\mu_i}\right)\delta^{(3)}(\vep-\vep')+\hat{V}_{ii}[\vep,\vep'],
$$
where $m_{i1}$ and $m_{i2}$ are the meson masses in the $i$-th channel, with the reduced mass
$$
\mu_i=\frac{m_{i1}m_{i2}}{m_{i1}+m_{i2}}.
$$
We use hats for the operators and square brackets for their arguments in order to distinguish between operators and the corresponding $c$-number functions. In particular:
\begin{eqnarray*}
&&\hat{\cal O}[\vep]c\equiv {\cal O}(\vep)c,\\
&&\hat{\cal O}[\vep]\chi(\vep)\equiv \int {\cal O}(\vep)\chi(\vep)d^3p ,\\
&&\hat{\cal O}[\vep,\vep']\chi(\vep)\equiv \int {\cal O}(\vep,\vep')\chi(\vep')d^3p'.
\end{eqnarray*}

In what follows we define the energy $E$ relative to the lowest threshold:
$$
M=m_{11}+m_{12}+E,\quad M_0=m_{11}+m_{12}+E_0.
$$

Then the mass parameters
\be
\Delta_i=(m_{i1}+m_{i2})-(m_{11}+m_{12}),
\label{Da}
\ee
define splittings between mesonic thresholds.

\subsection{$t$-matrix}

The system of coupled Lippmann--Schwinger equations for the various components of the off-shell $t$-matrix reads:
\be
\left\{
\begin{array}{l}
\ds t_{00}(E)=-\sum_k\int f_k(\veq)S_k(\veq)t_{k0}(\veq,E) d^3q,\\
\ds t_{i0}(\vep,E)=f_i(\vep)-\frac{f_i(\vep)t_{00}(E)}{E_0-E-i0}\\
\ds \qquad\quad-\sum_k\int V_{ik}(\vep,\veq)S_k(\veq)t_{k0}(\veq,E)d^3q\\
\ds t_{0i}(\vep,E)=f_i(\vep)-\sum_k\int f_k(\veq)S_k(\veq)t_{ki}(\veq,\vep,E) d^3q,\\
\ds t_{ij}(\vep,\vep',E)=V_{ij}(\vep,\vep')-\frac{f_i(\vep)t_{0j}(\vep',E)}{E_0-E-i0}\\
\ds\qquad\qquad-\sum_k\int V_{ik}(\vep,\veq)S_k(\veq)t_{kj}(\veq,\vep',E) d^3q,
\end{array}
\right.
\label{t0}
\ee
where
\be
S_i(\vep)=\frac{1}{p^2/(2\mu_i)-E+\Delta_i-i0}
\label{Ss}
\ee
is the propagators of the meson pair in the $i$-th channel.
Solution of system (\ref{t0}) can be found in the form (see also
Ref.~\cite{interplay}):
\bea
&&t_{00}(E)=-\frac{(E-E_0){\cal G}(E)}{E-E_0+{\cal G}(E)},\label{t00sol}\\
&&t_{0i}(\vep,E)=\frac{E-E_0}{E-E_0+{\cal G}(E)}\bar{\phi}_i(\vep),\label{t0isol}\\
&&t_{i0}(\vep,E)=\frac{E-E_0}{E-E_0+{\cal G}(E)}\phi_i(\vep),\label{ti0sol}\\
&&t_{ij}(\vep,\vep',E)=t_{ij}^V(\vep,\vep')+\frac{\phi_i(\vep)\bar{\phi}_j(\vep')}{E-E_0+{\cal G}(E)},\label{tijsol}
\eea
where
\bea
\ds\phi_i(\vep)&=&f_i(\vep)-\sum_k\int t_{ik}^V(\vep,\veq)S_k(\veq)f_k(\veq)d^3q,\label{sol11}\\
\ds\bar{\phi}_i(\vep')&=&f_i(\vep)-\sum_k\int S_k(\veq)f_k(\veq)t_{ki}^V(\veq,\vep')d^3q,
\label{sol1}\\
{\cal G}(E)&=&\sum_i\int f_i^2(\veq)S_i(\veq)d^3q\label{Gsol}\\
&-&\sum_{i,j}\int f_i(\vek)S_i(\vek)t_{ij}^V(\vek,\veq)S_j(\veq)f_j(\veq)d^3k d^3q\nonumber,
\eea
and the components of the $t$-matrix $t^V(\vep,\vep')$ by definition satisfy the last equation in system (\ref{t0}) with $f_i(\vep)=0$, that is,
\be
t_{ij}^V(\vep,\vep')=V_{ij}(\vep,\vep')-\sum_k\int V_{ik}(\vep,\veq)S_k(\veq)t_{kj}^V(\veq,\vep') d^3q,
\label{tVdef}
\ee
and thus the $t$-matrix $t^V(\vep,\vep')$ absorbs all the details of the direct interactions between mesonic channels.
It should be stressed that the procedure used to arrive at the solution (\ref{t00sol})--(\ref{tijsol})
automatically ensures unitarity, since it is based on solving a Schr\"odinger-type equation.

Finally, for further convenience, we define quantities $R_i$ and $R_i'$ such that
\begin{eqnarray*}
&&\int f_i^2(\veq)S_i(\veq)d^3q=f_{0i}^2(R_i+iI_i),\\
&&\int f_i(\veq)S_i(\veq)d^3q=f_{0i}(R'_i+iI_i),
\end{eqnarray*}
where $f_{0i}\equiv f_i(0)$ and
$$
I_i=4\pi^2\mu_ik_i,
$$
with the momenta $k_i$ defined similarly to Eq.~(\ref{ks0}), that is
\bea
k_i(E)&=&\sqrt{2\mu_i (E-\Delta_i)}\Theta(E-\Delta_i)\label{ks}\\
&+&i\sqrt{2\mu_i(\Delta_i-E)}\Theta(\Delta_i-E).\nonumber
\eea

\subsection{Bound states and continuum spectrum}

The Schr{\"o}dinger-like equation (\ref{Sheq}) admits both bound-state solutions and continuum spectrum.
If a bound state is present in the spectrum of the Hamiltonian (\ref{H}), the corresponding solution of Eq.~(\ref{Sheq}) takes the form:
\be
c_B=\cos\theta,\quad\chi_{iB}=S_i(\vep)\phi_i(\vep)\cos\theta,
\label{sol2}
\ee
where the hadronic Green's function $S_i(\vep)$ and the function $\phi_i(\vep)$ are defined in Eqs.~(\ref{Ss}) and (\ref{sol11}), respectively. The value of the angle $\theta$ follows from the normalisation condition for the w.f. (\ref{state}) which, for the bound state solution, reads:
$$
c_B^2+\sum_i\int d^3p|\chi_{iB}(\vep)|^2=1,
$$
so that
$$
\tan^2\theta=\sum_i\int S^2_i(\vep)\phi^2_i(\vep)d^3p.
$$

The Weinberg $Z$-factor (see Ref.~\cite{Weinberg}), which measures the
probability to find the compact component in the physical w.f. and
therefore encodes information on the nature of the state, is then:
\be
Z=|c_B|^2=\cos^2\theta.
\label{Zdef}
\ee

The $t$-matrix of Eqs.~(\ref{t00sol})--(\ref{tijsol}) develops a pole at the
bound-state energy $\varepsilon$
(that is, at $M_B=m_{11}+m_{12}-\varepsilon$), if
\be
E_0=-\varepsilon+{\cal G}(-\varepsilon),
\label{veEq}
\ee
where the loop function ${\cal G}(E)$ is defined in Eq.~(\ref{Gsol}). Alternatively, Eq.~(\ref{veEq}) can be obtained directly from Eqs.~(\ref{Sheq}) and (\ref{sol2}).
The generalisation to the case of multiple bound states is trivial.

The Hamiltonian of Eq.~(\ref{H}) possesses also a continuum spectrum. The solution of the Schr{\"o}\-din\-ger-like equation (\ref{Sheq}) with free asymptotics in the hadronic channel $i$ takes the form:
\bea
\chi^{(i)}_{j;\vek_i}(\vep)&=&\delta_{ij}\delta^{(3)}(\vep-\vek_i)-S_j(\vep)
t_{ji}(\vep,\vek_i,E),
\label{chi0}\\
c^{(i)}_{\vek_i}(E)&=&-\frac{t_{0i}(\vek_i,E)}{E_0-E},
\label{c}
\eea
where $k_i=\sqrt{2\mu_i(E-\Delta_i)}$.
Equation (\ref{c}) allows one to calculate the spectral density $w(E)$,
\be
w(E)=\sum_i \mu_i k_i\Theta(E-\Delta_i)\int \left|c^{(i)}_{\vek_i}(E)\right|^2d o_{\vek_i},
\label{wE}
\ee
which gives the probability to find the bare state in the continuum w.f. with
the given energy $E$ \cite{bhm}.
Then the overall normalisation condition reads:
$$
Z+\int_0^\infty w(E)dE=1,
$$
where $Z$ is defined in Eq.~(\ref{Zdef}). In case of existence of several bound states, the latter equation is obviously generalised as
$$
\sum_n Z_n+\int_{0}^\infty w(E)dE=1,
$$
where the corresponding $Z$-factors are given by expressions similar to Eq.~(\ref{Zdef}).

\section{Two-chanel case}\label{twochan}

In this chapter we present the simplest nontrivial application of the general formulae (\ref{t00sol})--(\ref{tijsol}) derived above, namely, we consider the two-channel solution and perform its low-energy reduction. We
have $\Delta_1=0$ and $\Delta_2\equiv\Delta>0$. In what follows, assuming
smallness of $\Delta$ as compared to the masses, we set $\mu_1=\mu_2=\mu$.

\subsection{$t$-matrix, bound states, and $Z$-factors}

In the scattering length approximation the 2x2 direct interaction $t$-matrix $t^V(\vep,\vep')$ (see
Eq.~(\ref{tVdef}) for its general definition) can be pa\-ra\-met\-ris\-ed in the form:
\be
t^V=\frac{1}{\mbox{Det}}
\left(
\begin{array}{cc}
\frac12(R_s+R_t)+iI_2&\frac12(R_t-R_s)\\
\frac12(R_t-R_s)&\frac12(R_s+R_t)+iI_1
\end{array}
\right),
\label{tv}
\ee
where
\be
\mbox{Det}=(R_sR_t-I_1I_2)+\frac{i}{2}(R_s+R_t)(I_1+I_2),
\label{detdef}
\ee
and the quantities $R_s$ and $R_t$ can be related to the (inverse) scattering
lengths in the singlet and triplet channels, respectively:
\be
R_s=4\pi^2\mu\gamma_s,\quad R_t=4\pi^2\mu\gamma_t.
\label{RsRt}
\ee
Note, those are not the physical scattering lengths, but only the
scattering lengths that would have been realised, if only the potential scattering
had been present; in general the physical scattering lengths are combinations
of $\gamma_s$, $\gamma_t$ and the pole parameters.  In what follows we stick to $\gamma_s$ and
$\gamma_t$ only as to convenient quantities parametrising the strength of the direct
interaction $t$-matrix $t^V$.

In what follows, we assume that the quark state is an isosinglet, setting
$$
R_1=R_2=R,\quad R_1'=R_2'=R',\quad f_{01}=f_{02}=f_0/\sqrt{2},
$$
and introduce three physical parameters, $E_f$ (Flatt{\'e} energy), $E_C$ (zero of the singlet $t$-matrix --- see below as well as Ref.~\cite{interplay}), and $g_f$ (Flatt{\'e} coupling constant) instead of the three bare parameters $E_0$, $R$, and $R'$:
\bea
&&E_f=E_0-\frac{f_0^2}{R_s}(RR_s-R'^2),\label{Ef1}\\
&&E_C=E_0-f_0^2(R_s+R-2R'),\label{EC1}\\
&&g_f=\frac{8\pi^2\mu}{R_s^2}f_0^2(R_s-R')^2.\label{gf1}
\eea

Then, after straightforward calculations, one can find for the various components of the $t$-matrix:
\bea
t_{11}&=&\frac{1}{8\pi^2\mu}\frac{\gamma_s(E-E_f)+(E-E_C)\left(\gamma_t+2ik_2\right)}{D(E)},\label{t11}\\
t_{12}&=&t_{21}=\frac{1}{8\pi^2\mu}\frac{\gamma_t(E-E_C)-\gamma_s(E-E_f)}{D(E)},\label{t12}\\
t_{22}&=&\frac{1}{8\pi^2\mu}\frac{\gamma_s(E-E_f)+(E-E_C)\left(\gamma_t+2ik_1\right)}{D(E)},\label{t22}
\eea
where
\bea
D(E)&=&\gamma_s\left(\gamma_t+\frac{i}{2}(k_1+k_2)\right)(E-E_f)\nonumber\\
&-&\left(k_1k_2-\frac{i}{2}\gamma_t(k_1+k_2)\right)(E-E_C).\label{Den}
\eea

Expressions (\ref{t11})--(\ref{t22}) coincide with those derived in Ref.~\cite{braaten2} if the parameters $\gamma_0$, $\gamma_1$, and $g$ used in Ref.~\cite{braaten2} are mapped to our parameters as
$$
\gamma_0=\gamma_s,\quad\gamma_1=\gamma_t,\quad g^2=\frac12 g_f.
$$
Notice also that a different definition of the isospin states is used in Ref.~\cite{braaten2} (see Eq.~(\ref{Cpar}) below), so that the $t$-matrix components $t_{11}$ and $t_{22}$ appear in Ref.~\cite{braaten2} with the opposite signs as compared to our Eqs.~(\ref{t11}) and (\ref{t22}).

It is instructive to construct the isospin singlet and triplet combinations,
which read:
\bea
t_s&=&\frac{1}{8\pi^2\mu}\frac{(E-E_C)\left(2\gamma_t+i(k_1+k_2)\right)}{D(E)},\label{ts}\\
t_t&=&\frac{1}{8\pi^2\mu}\frac{2\gamma_s(E-E_f)+i(k_1+k_2)(E-E_C)}{D(E)},\label{tt}\\
t_{st}&=&\frac{1}{8\pi^2\mu}\frac{i(k_2-k_1)(E-E_C)}{D(E)},\label{tst}
\eea
where, similarly to the single-channel case \cite{interplay}, the
$t$-matrix $t_s$ possesses a zero at $E=E_C$. In addition, expressions
(\ref{ts})--(\ref{tst}) contain explicitly the leading isospin violation:
$k_1\neq k_2$ allows for transitions between isospin singlet and triplet states.

For the bound-state solution of the Schr{\" o}dinger-like equation (\ref{Sheq}) the $Z$-factor can be found explicitly in the form:
\be
Z=\frac{1}{1+{\cal G}'(-\varepsilon)}=\frac{1}{1+\tan^2\theta},
\label{Z2}
\ee
with
\bea
&&\tan^2\theta=\frac{\mu(E_C+\varepsilon)}{\kappa_1\kappa_2}\label{tg}\\
&&\times\frac{\kappa_1(\gamma_t-\kappa_1)^2+\kappa_2(\gamma_t-\kappa_2)^2}{(\kappa_1+\kappa_2-2\gamma_t)\left(\gamma_s\gamma_t+\kappa_1\kappa_2-\frac12(\gamma_s+\gamma_t)(\kappa_1+\kappa_2)\right)},\nonumber
\eea
where $\kappa_i\equiv\mbox{Im}(k_i(-\varepsilon))$ --- see Eq.~(\ref{ks}) for
the general case.

\subsection{Limiting cases}

In this chapter we check various limiting cases of the low-energy two-channel formulae derived above in this Section.

We start from the single-channel case. In order to approach the single-channel limit one is to set $\Delta=0$ everywhere which, in particular, implies the substitution
$$
k_1(E)=k_2(E)=k(E).
$$

Then one readily finds for the singlet $t$-matrix and for the $Z$-factor:
\be
t_s^{-1}(E)=4\pi^2\mu \left(\frac{E-E_f}{E-E_C}\gamma_s+ik\right),
\label{tmin1}
\ee
$$
\frac{1-Z}{Z}=\tan^2\theta=\frac{E_C+\varepsilon}{2\varepsilon}\frac{\sqrt{2\mu\varepsilon}}{\sqrt{2\mu\varepsilon}
-\gamma_s},
$$
respectively, which coincide with the corresponding expressions derived in Ref.~\cite{interplay}, after an obvious identification $\gamma_s=\gamma_V$.

The zero $E_C$ affects the dynamics of the system, if it appears in the
near-threshold region. For the coupled channel situation at hand this happens for
$|E_C|\lesssim\Delta$ --- c.f. the discussion in
Ref.~\cite{interplay} for the single-channel case.  The value of $E_C$ is, by
construction, controlled by the singlet inverse scattering length
$\gamma_s$, since it follows from Eqs.~(\ref{Ef1})--(\ref{gf1}) that
 \be
E_C=E_f-\frac12 g_f\gamma_s.
\label{ECEf}
\ee
Thus the zero $E_C$ is generated far away from the near-threshold region of interest, $|E_C|\gg\Delta$, if
\be
|\gamma_s|\gg \frac{\Delta}{g_f}.
\label{gslim}
\ee
For this estimate we assumed $|E_f|\lesssim \Delta$.
We therefore conclude that, if Eq.~(\ref{gslim}) holds, then
$\gamma_s$ might be assumed to be infinite or, correspondingly, that the
singlet scattering length is negligibly small. It should be stressed however
that this does not imply the absence of meson--meson interaction at all in
the singlet channel. Indeed, as explained before (see the discussion below Eq.~(\ref{RsRt})), the scattering lengths $\gamma_s$ and $\gamma_t$ are only residual scattering lengths.
Similarly to the single-channel case we expect the line shapes and the interpretation
thereof to simplify tremendously for large values of the $E_C$, however, as we
shall see, additional structures can be introduced through the triplet
channel.

Indeed, the momenta $k_1$
and $k_2$ enter expressions for the $t$-matrix not in the form of a naive sum
$k_1+k_2$, but in an entangled way, including the product $k_1k_2$ (see
also Ref.~\cite{braaten2} for the discussion of the related effect on the line
shapes for the $X(3872)$). This entanglement of mesonic channels is governed
by the triplet inverse scattering lengths $\gamma_t$ and it becomes strong for
$|\gamma_t|\lesssim\sqrt{\mu\Delta}$.  Therefore, the large-$\gamma_t$ limit
is achieved for
\be
|\gamma_t|\gg\sqrt{\mu\Delta}
\label{gtlim}
\ee
in the general two-channel formulae.

We therefore identify four possible cases:
\begin{itemize}
\item Case (i): $|\gamma_s|\to\infty$ and $|\gamma_t|\to\infty$.
\item Case (ii): small $\gamma_s$ and $|\gamma_t|\to\infty$.
\item Case (iii): $|\gamma_s|\to\infty$ and small $\gamma_t$.
\item Case (iv): both $\gamma_s$ and $\gamma_t$ are small.
\end{itemize}

To approach Case (i) above one is to assume both conditions (\ref{gslim}) and
(\ref{gtlim}) to hold simultaneously. In this case one arrives readily at the
two-channel Flatt{\'e} denominator which contains trivially the sum of the two
channels, that is, at Eq.~(\ref{Dsimple}). For the Weinberg $Z$ factor we find:
$$
Z=\left(1+\frac{\mu g_f (\kappa_1+\kappa_2)}{4\kappa_1\kappa_2}\right)^{-1},
$$
with the standard pattern (c.f. Ref.~\cite{evi}), namely for $g_f\to 0$ we have $Z\to 1$, which
corresponds to a compact state, while for $g_f\to \infty$ we have $Z\to 0$
and the state is to be interpreted as dominantly molecular\footnote{It is
important to stress that from the Flatt{\'e} parametrisation alone it is
not possible to reliably extract $g_f$, if it is large, due to a scale
invariance of the parametrisation in that limit~\cite{ourflatte}.}.

Case (ii) implies that, while entanglement of the meso\-nic channels still
remains irrelevant, the zero $E_C$ starts to approach the near-threshold
region. Then, in this limit, one arrives immediately at the trivial
generalisation of the single-channel expressions found in
Ref.~\cite{interplay}. In particular, the Flatt{\'e} denominator takes the form:
\be
{\cal D}_F(E)=E-E_f-\frac{(E-E_f)^2}{E-E_C}+\frac{i}{4}g_f(k_1+k_2),
\label{D2}
\ee
which simply amounts to the substitution $k\to\frac12(k_1+k_2)$ in the
corresponding single-channel formula. Then the entire analysis performed in
Ref.~\cite{interplay} with respect to the role played by the zero $E_C$ for
the dynamics of the system applies directly to the two-channel case at
hand. In order to keep in Eq.~(\ref{D2}) the first non-vanishing correction in
the $1/\gamma_t$ expansion one is to substitute:
$$
k_1+k_2\to k_1+k_2-\frac{i}{2\gamma_t}(k_1-k_2)^2.
$$

In Case (iii), the zero $E_C$ plays no role in the dynamics of the system, the
new effect as compared to the simplest Case (i) being the entanglement of the
mesonic channels governed by the triplet (inverse) scattering length
$\gamma_t$. The corresponding expressions for various components of the
$t$-matrix read then:
\bea
t_s&=&\frac{g_f}{8\pi^2\mu}\frac{1}{{\cal D}(E)},\\ t_t&=&\frac{E-E_f+\frac{i}{4}g_f(k_1+k_2)}{4\pi^2\mu(\gamma_t+\frac{i}{2}(k_1+k_2))}\frac{1}{{\cal D}(E)},\\ t_{st}&=&\frac{ig_f}{16\pi^2\mu}\frac{k_2-k_1}{\gamma_t+\frac{i}{2}(k_1+k_2)}\frac{1}{{\cal D}(E)},
\eea
where
$$
{\cal D}(E)=E-E_f+\frac{i}{4}g_f(k_1+k_2)+\frac{1}{4}g_f\frac{(k_1-k_2)^2}{2\gamma_t+i(k_1+k_2)}.
$$
The effect of the channels entanglement for the $X(3872)$ resonance was
discussed in Ref.~\cite{braaten2}.  As a result new structures in the
corresponding line shapes were observed especially far away from the threshold, at
$E\gg\Delta$. In this paper we choose to stay within the range of
applicability of the effective-range expansion used to derive all formulae
above, that is in the near-threshold region $|E|\lesssim\Delta$.

Finally, in Case (iv) one cannot simplify the general formulae derived in the
previous chapter and is forced to face the full complexity of the interplay of
quark and mesonic degrees of freedom as well as that of the mesonic channels
entanglement.

One should have in mind that, in reality, inelastic channels with far-away
thresholds also exist, and could be dealt with the help of the general
multi-channel formalism of Sec.~\ref{Mcp}. However, these remote thresholds do
not produce extra energy (momentum)-dependent contributions to the
near-threshold amplitude (though their presence in the system may modify the
line shapes appreciably --- see, for example, Ref.~\cite{ourX2}). The net
effect of such thresholds can be taken into account by including imaginary
parts of inverse scattering lengths $\gamma_s$ and $\gamma_t$ (if the
inelastic channels are coupled to hadronic components of the w.f.) and
Flatt{\'e} energy $E_f$ (if the inelastic channels are coupled to the quark
component). Since the present research is aimed solely at the discussion of
multi-channel dynamics in a near-threshold resonance, we regard the investigations
of remote thresholds as lying beyond the scope of this paper.

\section{An illustrative example}\label{example}

In previous chapters we derived general expressions for the w.f.'s and
$t$-matrices for a near-threshold resonance which incorporated effects due to
quark and multi-channel mesonic dynamics. For the two-channel case we performed the low-energy reduction of the solution found and arrived
at rather simple expressions. We then pinpointed parameters which governed the
dynamics of the system, in particular, the appearance of the zero $E_C$ in the
near-threshold region and the entanglement of mesonic channels. This finalised
a purely theoretical analysis of the interplay of quark and meson degrees of
freedom in a near-threshold resonance in the multi-channel case. In this
chapter, we proceed with phenomenological applications of the general formulae
derived above. In particular, we study line shapes of the $X$ production under various assumptions concerning the
production mechanism. Keeping in mind that the charmonium state
$X(3872)$ provides a paradigmatic example of a two-channel situation, we use it as an anchor which helps us to
fix several sets of parameters. We caution the readers however of a naive identification of these sets of parameters, used solely for illustrative purposes, with real fits to the experimental data for the state $X(3872)$. As was explained before, the analysis of the experimental data for the $X(3872)$ lies beyond the scope of the present paper, since the aim of the latter is developing and improving a phenomenological tool which could be used then for the analysis of experimenal data on various near-threshold resonances. Three more comments are in order here before we proceed.

First, we fix the quantum numbers of the $X$ to be $1^{++}$ which is compatible with the $X(3872)$ state. Let us remind the readers that the problem of the quantum numbers for the $X(3872)$ is not fully resolved yet: while the analyses of the $J/\psi \pi^+\pi^-$ decay mode of the $X(3872)$ yields either $1^{++}$ or $2^{-+}$ quantum numbers \cite{rho}, the recent analysis of the $J/\psi \pi^+\pi^-\pi^0$ mode seems to favour the $2^{-+}$ assignment \cite{omega}, though the $1^{++}$ option is not excluded. As shown in Ref.~\cite{1D2}, the $X(3872)$ cannot be a naive $c \bar c$ $2^{-+}$ state and, were the $2^{-+}$ quantum numbers confirmed, very exotic explanations for the $X(3872)$ would have to be invoked. In the absence of such a confirmation we stick to the most conventional $1^{++}$ assignment for the $X(3872)$. With the given quantum numbers, the lowest open-charm channels are $D\bar{D}^*$\footnote{An obvious shorthand
notation is used here and in what follows: $D \bar D^* \equiv
\frac{1}{\sqrt{2}}(D \bar D^* + \bar D D^*)$ (see Eq.~(\ref{Cpar})).} in the relative $S$-wave, and
the $X(3872)$ is known to be located very close to the $D^0\bar{D}^{*0}$
threshold. Identification of the channels is obvious in this case: channel 1
is associated with the neutral component $D^0\bar{D}^{*0}$ while channel 2 ---
with the charged component $D^+\bar{D}^{*-}$.

As the second comment we would like to note that, for the definition of the $C$-even(odd) state, we stick to the convention of Ref.~\cite{Cparity} for the $C$-parity operator $\hat{C}$ and $C$-parity eigenstates:
\be
\bar{P}(\bar{V})=\hat{C}P(V),\quad\hat{C}|P\bar{V}\pm\bar{P}V\rangle=\pm|P\bar{V}\pm\bar{P}V\rangle,
\label{Cpar}
\ee
with $P$ and $V$ denoting the pseudoscalar and the vector, respectively. Further details of the ambiguity
in the definition of $C$-eigenvalue states can be found in Ref.~\cite{Cparity}.

The third comment concerns the stability of the $X$ constituents $M_{11}$, $M_{12}$, $M_{21}$, and $M_{22}$ (see Eq.~(\ref{state}) above). In order to get an insight into this problem we resort again to the example of the charmonium $X(3872)$. The $D^*$ mesons are unstable, and the question is whether the account for the $D^*$ finite width is important. On one hand, if the $X(3872)$ is a bound state and
thus its pole is located on the first sheet below the $D^0\bar{D}^{*0}$ threshold,
then finite width effects produce a spectacular below-threshold peak in the
$D^0 \bar{D}^0 \pi^0$ mass distribution --- see Refs.~\cite{ourX2,braatenfw}.
As argued in Ref.~\cite{ourX2}, observation or non-observation of this
below-threshold peak is a potentially powerful tool to discriminate between
various models for the $X(3872)$. On the other hand, the existing data analyses
does not allow one to exploit this tool. As stressed in
Refs.~\cite{braatenfw}, the analyses procedure is biased by the
assumption that all $D^0 \bar{D}^0 \pi^0$ events come from the
$D^0\bar{D}^{*0}$ mode, so that events from below the $D^0\bar{D}^{*0}$
threshold are move above it. In addition, the existing experimental resolution is too
coarse to investigate in detail the narrow region around the $D^0\bar{D}^{*0}$
threshold. Finally, the finite width effects are known to be negligible above threshold
for the energies larger than the width of the $D^{*0}$, that is, larger than
approximately 100~keV (see the discussions in
Refs.~\cite{ourX2,braatenfw,finwidth}). Therefore in what follows we neglect these
effects.

We stick to the neutral mesons in the final state and consider production of
the $X$ through the hadronic as well as the quark components. The
corresponding amplitudes read:
\bea
{\cal M}_{h_1}&=&{\cal F}_{h_1}-{\cal F}_{h_1}\int S_1(\vep)t_{11}(\vep,\vek,E)d^3p,\label{h1}\\
{\cal M}_{h_2}&=&{\cal F}_{h_2}-{\cal F}_{h_2}\int S_2(\vep)t_{21}(\vep,\vek,E)d^3p,\label{h2}\\
{\cal M}_q&=&-{\cal F}_q\frac{1}{E-E_0}t_{01}(E),
\eea
where $E=k^2/(2\mu)$, and ${\cal F}$'s denote point-like sources which incorporate the details of the short-ranged
dynamics of the $X$ production. In what follows we neglect Born terms
in the production through the hadronic components (the first term on the r.h.s. of Eqs.~(\ref{h1}) and (\ref{h2})). This is justified if the $t$-matrix elements have a near-threshold pole (a detailed discussion can be found in Ref.~\cite{interplay}).

Then, if one neglects the interference between three possible production mechanisms, it is straightforward to find the corresponding branchings:
\bea
&&\frac{dBr_q}{dE}=\Theta(E)\frac{{\cal B}_0\sqrt{E}}{|D(E)|^2}{\gamma_s}^2\left|\gamma_t+ik_2\right|^2,\label{BrqX}\\ &&\frac{dBr_{h_1}}{dE}=
\Theta(E)\frac{{\cal B}_1\sqrt{E}}{|D(E)|^2}\left|\vphantom{\int_0^1}\gamma_s\left(E-E_f\right)\right.\label{Brh1X}\\ &&\hspace*{29mm}+\left.(\gamma_t+2ik_2)
\left(E-E_f+\frac12g_f\gamma_s\right)\right|^2,\nonumber\\ &&\frac{dBr_{h_2}}{dE}=
\Theta(E)\frac{{\cal B}_2\sqrt{E}}{|D(E)|^2}\left|\vphantom{\int_0^1}\gamma_s\left(E-E_f\right)\right.\label{Brh2X}\\ &&\hspace*{43mm}\left.-\gamma_t\left(E-E_f+\frac12g_f\gamma_s\right)\right|^2,\nonumber
\eea
where the denominator $D(E)$ is given by the following expression (see Eqs.~(\ref{Den}) and (\ref{ECEf})):
\bea
&&D(E)=\gamma_s\left(\gamma_t+\frac{i}{2}(k_1+k_2)\right)\left(E-E_f\right)
\label{Den2}\\
&&-\left(k_1k_2-\frac{i}{2}\gamma_t(k_1+k_2)\right)\left(E-E_f+\frac12 g_f\gamma_s\right).\nonumber
\eea
The coefficients ${\cal B}_0$, ${\cal B}_1$, and ${\cal B}_2$ are proportional to the short-range factors
$|{\cal F}_q|^2$, $|{\cal F}_{h1}|^2$, and $|{\cal F}_{h2}|^2$ and, in the case of production via hadronic components, include the effects of renormalisation of loop functions in Eqs.~(\ref{h1}) and (\ref{h2}).
The formulae for the production rates can be obtained from the general ones given in Ref.~\cite{braaten2}, if one assumes the dominance of certain production mechanism, and neglects the finite width effects.

In order to exemplify the line shapes given by Eqs. (\ref{BrqX})--(\ref{Brh2X}) we fix parameters as
\be
\mu=966.5~\mbox{MeV},\quad\Delta=8.1~\mbox{MeV},\quad g_f=0.25
\label{set}
\ee
and consider Cases (i)--(iv) introduced before. Other parameters are quoted in Table~\ref{t1}. Notice that, for infinite inverse scattering lengths $\gamma_s$ and $\gamma_t$ their signs obviously play no role. Finite $\gamma_s$ and $\gamma_t$ are chosen to be small and negative, providing in such a way a strong attraction in the system which is, however, not strong enough to generate a bound state by itself. Extra attraction is provided by
the coupling to the quark state and, for each case, the Flatt{\'e} parameter $E_f$ is tuned to provide a bound state at $E_B=-\varepsilon=-0.5$~MeV.

\begin{center}
\begin{figure}[t]
\centerline{\epsfig{file=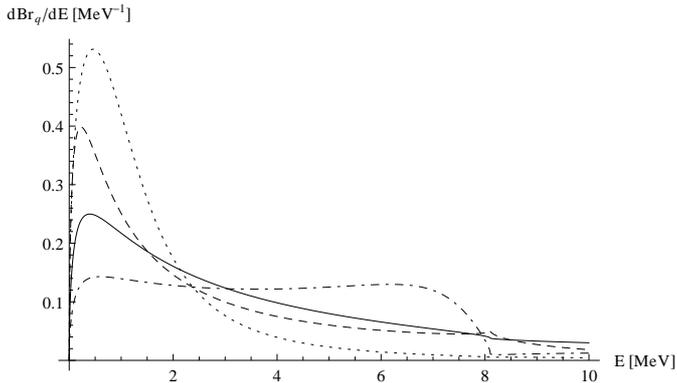, width=8.8cm}}
\caption{Production rate through the quark component (\ref{BrqX}). Cases (i)--(iv) are given by the solid, dashed, dashed-dotted, and dotted lines, respectively.}\label{brqFig}
\end{figure}
\end{center}

\begin{center}
\begin{figure}[t]
\centerline{\epsfig{file=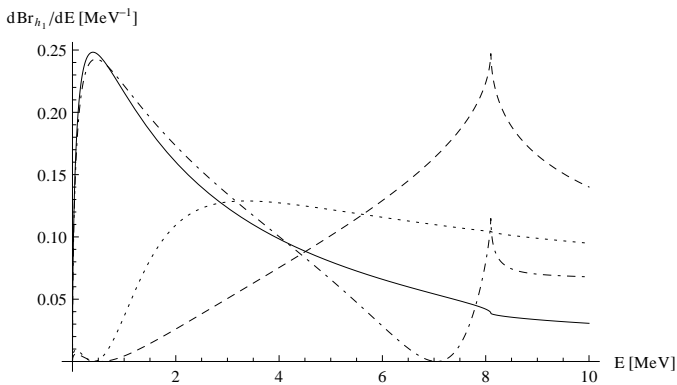, width=8.8cm}}
\caption{Production rate through the first hadronic component (\ref{Brh1X}). Cases (i)--(iv) are given by the solid, dashed, dashed-dotted, and dotted lines, respectively.}\label{brh1Fig}
\end{figure}
\end{center}

\begin{center}
\begin{figure}[t]
\centerline{\epsfig{file=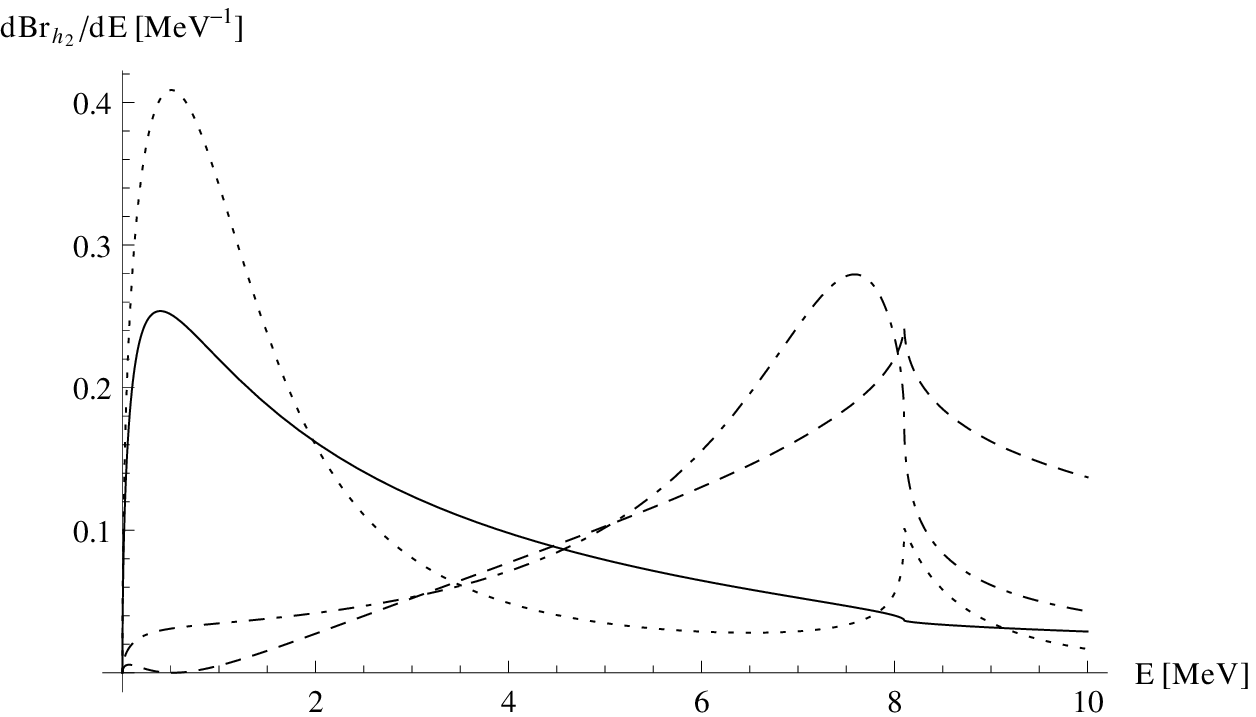, width=8.8cm}}
\caption{Production rate through the second hadronic component (\ref{Brh2X}). Cases (i)--(iv) are given by the solid, dashed, dashed-dotted, and dotted lines, respectively.}\label{brh2Fig}
\end{figure}
\end{center}

\begin{table}[t]
\begin{center}
\begin{tabular}{|c|c|c|c|c|c|}
\hline
&$\gamma_s$, MeV&$\gamma_t$, MeV&$E_f$, MeV&Line&Z\\
\hline
(i)&$\pm\infty$&$\pm\infty$&-10.47&solid&0.30\\
\hline
(ii)&-30&$\pm\infty$&-3.22&dashed&0.85\\
\hline
(iii)&$\pm\infty$&-30&-7.77&dashed&0.19\\
&&&&-dotted&\\
\hline
(iv)&-30&-30&-2.97&dotted&0.67\\
\hline
\end{tabular}
\end{center}
\caption{Scattering length parameters, the Flatt{\'e} parameter $E_f$, and
the $Z$-factor for the bound-state at $\varepsilon=-0.5$~MeV for Cases (i)--(iv).}
\label{t1}
\end{table}
In Figs.~\ref{brqFig}-\ref{brh2Fig} we plot the line shapes for the $X$
production through the quark component (see Eq.~(\ref{BrqX}) and
Fig.~\ref{brqFig}) and both hadronic components (see Eq.~(\ref{Brh1X}) and
Fig.~\ref{brh1Fig} as well as Eq.~(\ref{Brh2X}) and Fig.~\ref{brh2Fig},
respectively) for all four above Cases (i)--(iv). For each individual curve in
Figs.~\ref{brqFig}-\ref{brh2Fig} the integral over the near-threshold region
(chosen to be from 0 to 10~MeV) equals to unity (in the units of MeV$^{-1}$)
which fixes the overall factors ${\cal B}_0$, ${\cal B}_1$, and ${\cal B}_2$.

\begin{figure*}[t]
\begin{center}
\begin{tabular}{cccc}
\epsfig{file=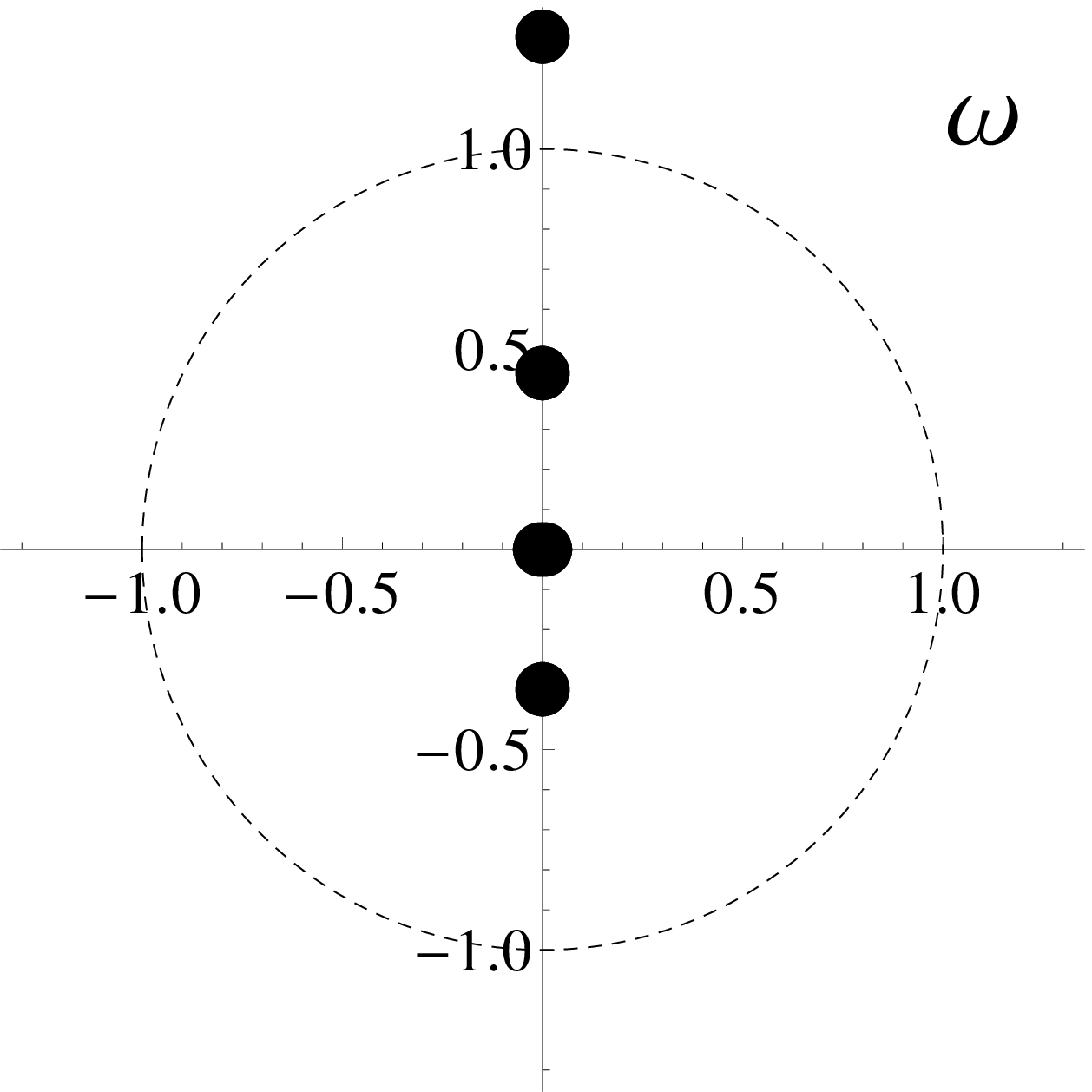, width=4cm}&\epsfig{file=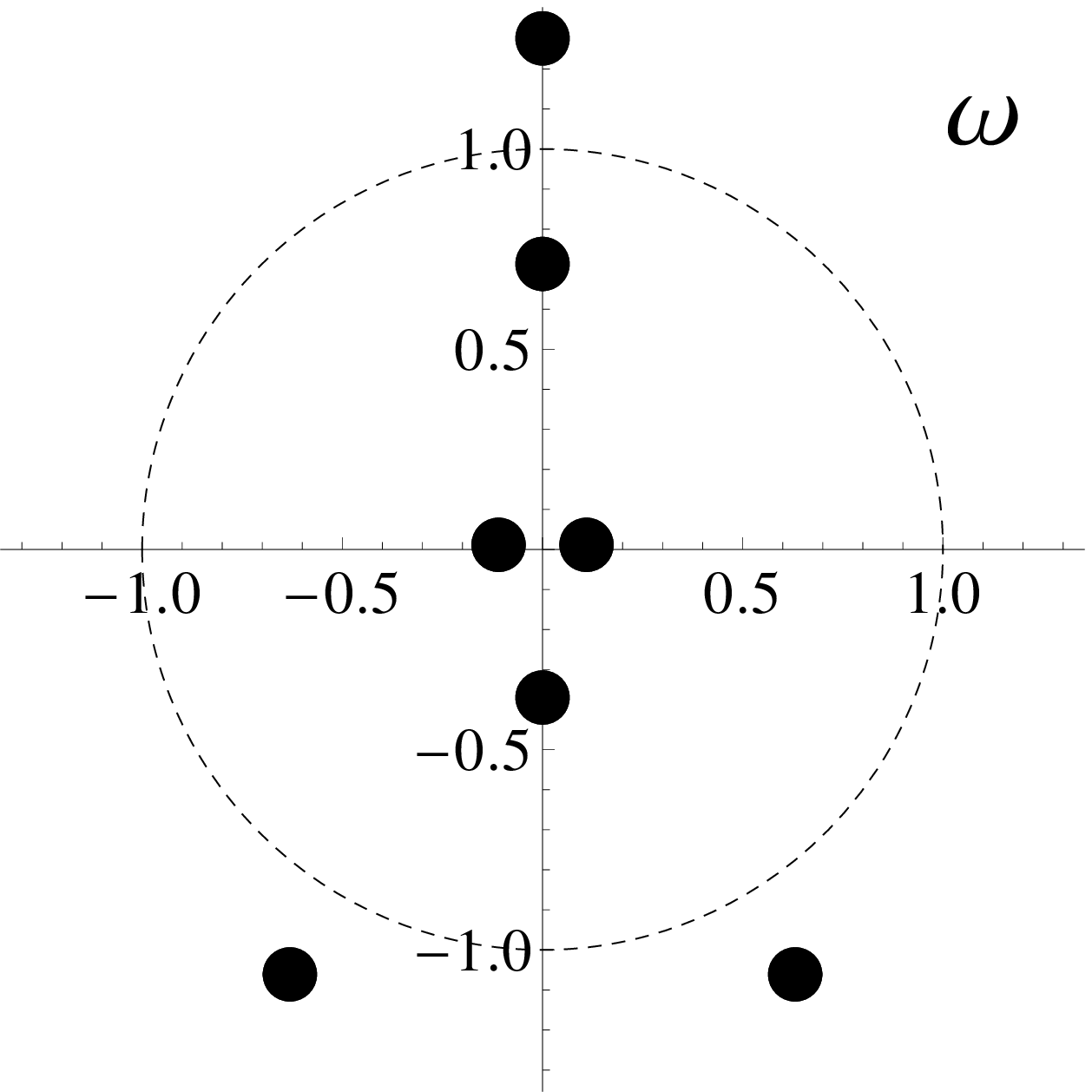, width=4cm}&
\epsfig{file=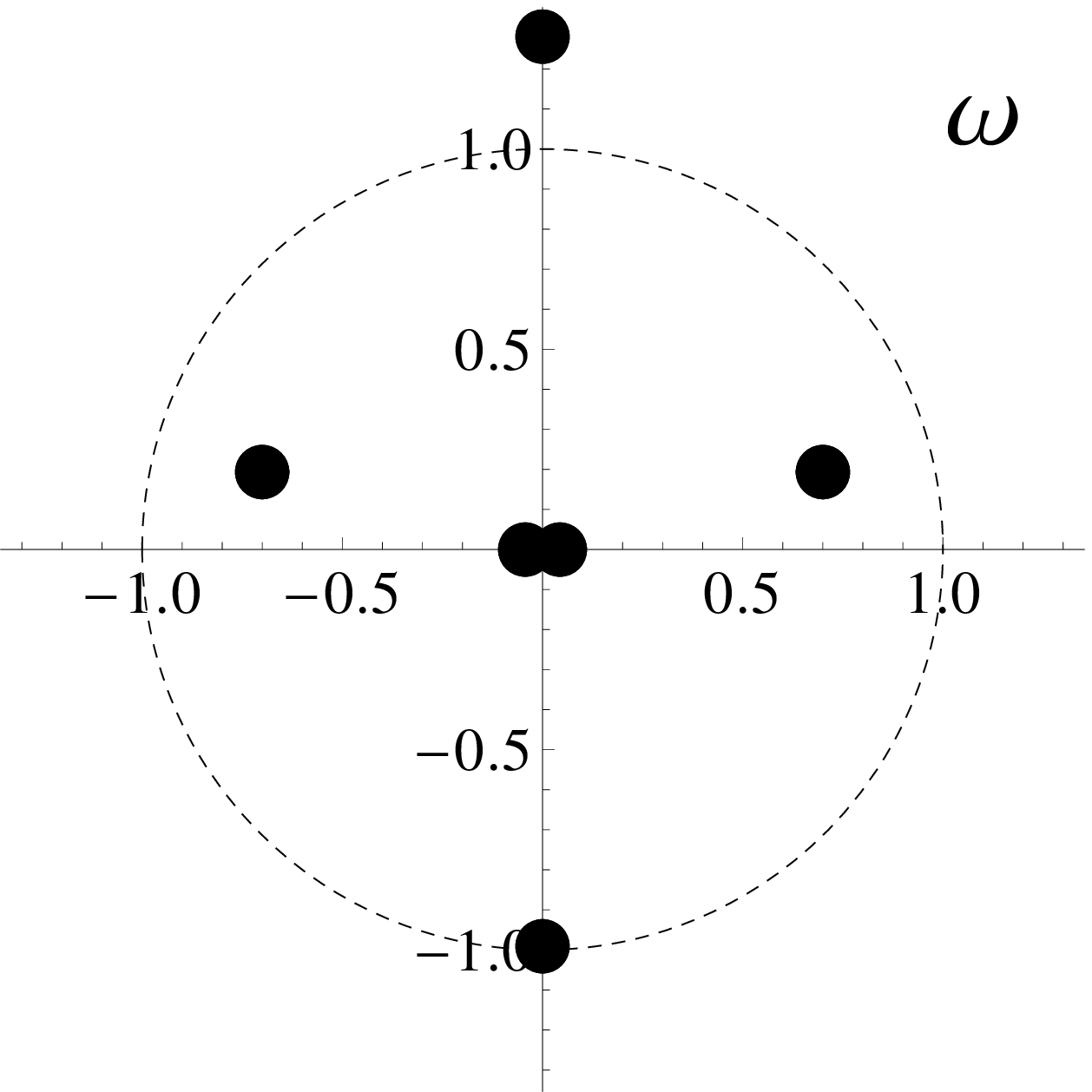, width=4cm}&\epsfig{file=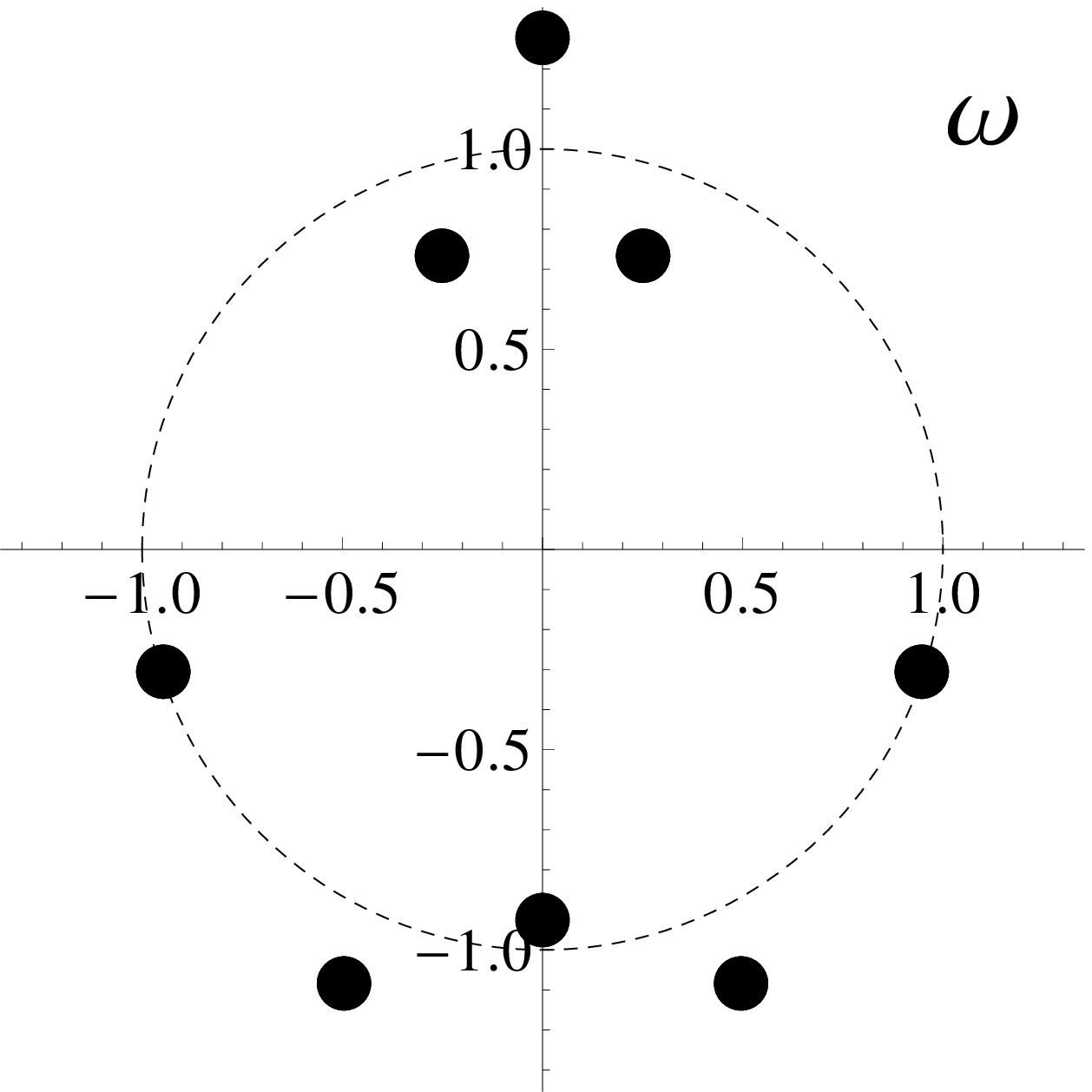, width=4cm}\\
Case (i)&Case (ii)& Case (iii)&Case (iv)
\end{tabular}
\caption{Near-threshold poles in the $\omega$-plane for Cases (i)--(iv).}\label{poles}
\end{center}
\end{figure*}

Finally, in Table~\ref{t2} and  Fig.~\ref{poles}, we show pole positions for all four Cases
(i)--(iv). To this end we
introduce the plane of the complex variable $\omega$ --- see
Appendix~\ref{omegaApp} for the definition of the variable $\omega$ and for the
mapping on the Riemann sheets.


\begin{table}[t]
\begin{center}
\begin{tabular}{|c|c|c|c|c|}
\hline
&I&II&III&IV\\
\hline
(i)&i1.28&i0.44&-i0.35&\\
&(-0.5)&(-6.78)&(-12.80)&\\
\hline
(ii)&i1.28&i0.71&&$\pm$ 0.63-i1.06\\
&(-0.5)&(-0.96)&&(1.94$\mp$ i1.54)\\
\hline
(iii)&i1.28&$\pm$0.70+i0.19&-i0.99&\\
&(-0.5)&(8.26$\mp$i1.42)&(-6.6$\times10^{-4}$)&\\
\hline
(iv)&i1.28&$\pm$0.25+i0.73&$\pm$0.95-i0.31&$\pm$0.50-i1.08\\
&(-0.5)&(0.42$\mp$i1.32)&(7.73$\pm$i0.02)&(1.24$\mp$i1.10)\\
&&&-i0.93&\\
&&&(-0.05)&\\
\hline
\end{tabular}
\end{center}
\caption{Positions of near-threshold poles in $\omega$-plane (in $E$-plane in MeV) for Cases (i)--(iv). The column labels I-IV mark the corresponding Riemann sheets. See also Fig.~\ref{poles} for the graphical presentation of the poles. Notice that some poles present in Fig.~\ref{poles} are omitted in the table since the corresponding energies lie far beyond the near-threshold region of interest.}
\label{t2}
\end{table}

A comment on the factorisation approximation used before in the production rates (\ref{BrqX})--(\ref{Brh2X}) is in order here.
For the $X(3872)$, this factorisation approximation for the $D^0 \bar D^{*0}$ production
via the hadronic component implies that in the charged $B$-meson decays, the $X$
is produced through the neutral component 1 while, in the neutral $B$-meson
decays, the $X$ is produced through the charged component 2. The $X$
production through the quark component 0 is, obviously, the same for charged
and neutral $B$-meson decays.

One could have naively concluded that, with data on the $X(3872)$ production
in both charged and neutral $B$-meson decays available, it is possible to
investigate the full complexity of quark and hadronic dynamics responsible for
the nature of the $X$ resonance. In reality, however, both quark and hadronic
mechanisms contribute and interfere and, with a suitable choice of the
production parameters ${\cal F}$'s, the resulting near-threshold line shapes
could be smooth enough even in the presence of competing quark and meson
dynamics --- see the examples given in Ref.~\cite{braaten2}. We emphasise
once again that the line shapes depicted in Figs.~\ref{brqFig}-\ref{brh2Fig},
where each figure originates from a particular production mechanisms and thus
interferences are neglected,
should not be viewed as  realistic and suitable for the
analysis of experimental data. With the sets of parameters and
assumptions used we more  aim to indicate the full complexity of near-threshold
phenomena coming from the interplay of the quark and meson degrees of freedom
as well as from the direct interaction between mesonic channels in an $X$-like
resonance. Notice that in the particular case of the $X(3872)$, the production
line shapes do not exhibit irregularities in the near-threshold region (up to
possible existence of a bound-state peak which, as discussed above, is not
observable due to the current situation with data analyses and the
experimental resolution).  The $X(3872)$ line shapes can be described
therefore by simple Flatt{\'e} formulae, which reduce to those of the scattering length
approximation, when the effective coupling constant is large (see
Refs.~\cite{ourflatte} for the details). Such a scenario is exemplified by our
Case (i) above: there is only one near-threshold pole corresponding to the
bound state, which is predominantly  molecular. On the contrary, our Cases
(ii)--(iv) yield several near-threshold poles in addition to the bound-state
one, and the line shapes are strongly distorted either by the $t$-matrix zero or by
the channel entanglement --- see Table~\ref{t2} and Fig.~\ref{poles}. Such
exotic scenarios seem to be largely excluded by the existing data on the
$X(3872)$.

\section{Summary}\label{sum}

In the multi-channel situation the dynamics of the resonance is affected by
both the interplay of the quark and meson dynamics as well as by the direct
interactions between mesonic channels. As seen from the examples given above
both these effects may give rise to quite peculiar properties of the resonance
line shapes. Thus one concludes that, if the experimental data exhibit such
properties, the resonance is generated by this complicated interplay. The
converse is not necessarily true: the interference between different
production mechanisms is possible, which could tame the resulting line
shapes. This, however, requires some fine tuning for the interference
between different production mechanisms.

Our studies allow one to outline a sensible procedure to investigate the
properties of near-threshold resonances: 
\begin{enumerate}
\item If data exhibit irregular behaviour in the near-thresh\-old region, the latter may be attributed to the interplay of the quark and meson degrees of freedom, and formulae for the general Case (iv) above or, possibly, for one of its limits, as given by Case (ii) or Case (iii), should be used in the analysis. 
\item If data in the near-threshold region does not exhibit any irregularities, simple Flatt{\'e} approximation --- see Eq.~(\ref{Dsimple}) --- is to be used. Indeed, although fine tuning of parameters in the general near-threshold formulae is possible in this case, which tames the resulting line shape, staying close to the threshold, it is not possible, as a matter of principle, to discriminate between different parameter sets which produce identical near-threshold behaviour. 
Then, in order to get insight into the nature of the resonance one is to analyse data either far from the threshold region or data for other production channels for the given resonance. In the former case, staying far from the threshold, one has to face the full complexity of the multi-channel formalism described in Sec.~\ref{Mcp} above. No model-independent analysis is possible in this case and one has to rely upon model-dependent assumptions in order to arrive at simpler expressions. 
\item If no data are available far from the threshold, the general formalism of Sec.~\ref{Mcp} can be used in order to study to what extend (statistics, resolution, binning procedure, and so on) the data would need to improve in order to get sensitive to the structures potentially present.
\end{enumerate}

No complicated structures are observed in the near-threshold region in the existing data on the $X(3872)$
production in $B$-meson decays. So, employing the analysis procedure outlined above, one concludes that, up to possible
interference effects, the $X(3872)$ line shapes in the near-threshold region
can be well described by simple Fla\-tt{\'e} formulae, like Eq.~(\ref{Dsimple}).

\begin{acknowledgement}
We acknowledge useful discussions with E.~Braaten.
The work was supported in parts by funds provided from the Helmholtz
Association (via grants VH-NG-222, VH-VI-231), by the DFG (via grants SFB/TR 16
and 436 RUS 113/991/0-1), by the EU HadronPhysics2 project, by the RFFI (via grants RFFI-09-02-91342-NNIOa and
RFFI-09-02-00629a). The work of Yu.K. and A.N. was supported by the State
Corporation of Russian Federation ``Rosatom''.
\end{acknowledgement}

\appendix

\section{The pole structure and the $\omega$-plane}\label{omegaApp}

The structure of the poles in the two-channel case can be visualised with the mapping of the four-sheeted Riemann surface onto the single complex plane \cite{kato}.
For a given energy $E$ one has:
\be
E=\frac{k_1^2}{2\mu_1},\quad E=\frac{k_2^2}{2\mu_2}+\Delta,
\label{Ek1k2}
\ee
so that, instead of two complex momenta $k_1$ and $k_2$ constrained by two conditions (\ref{Ek1k2}), it is convenient to introduce one complex variable $\omega$ defined such that
\be
k_1=\sqrt{\frac{\mu_1\Delta}{2}}\left(\omega+\frac{1}{\omega}\right),\quad
k_2=\sqrt{\frac{\mu_2\Delta}{2}}\left(\omega-\frac{1}{\omega}\right),
\label{omegak1k2}
\ee
which gives for the energy:
$$
E=\frac{\Delta}{4}\left(\omega^2+\frac{1}{\omega^2}+2\right).
$$
By construction, the complex $\omega$-plane is free of unitary cuts.

\begin{figure}[ht]
\begin{center}
\centerline{\epsfig{file=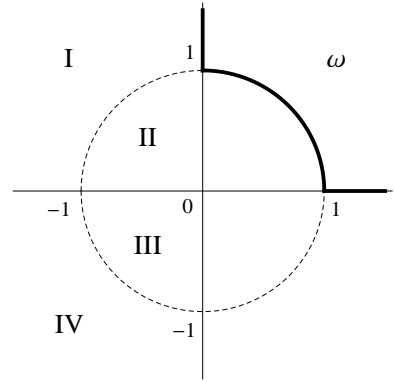, width=5cm}}
\caption{Riemann sheets in $\omega$-plane.}\label{omegaFig}
\end{center}
\end{figure}

In Fig.~\ref{omegaFig} we show the mapping of Riemann sheets onto the $\omega$ complex plane. Thick solid line corresponds to the real values of the energy $E$ on the first sheet. The sheets are labelled as follows:
\begin{eqnarray*}
{\rm I:}&&\quad{\rm Im}~k_1>0,\quad{\rm Im}~k_2>0,\\
{\rm II:}&&\quad{\rm Im}~k_1<0,\quad{\rm Im}~k_2>0,\\
{\rm III:}&&\quad{\rm Im}~k_1>0,\quad{\rm Im}~k_2<0,\\
{\rm IV:}&&\quad{\rm Im}~k_1<0,\quad{\rm Im}~k_2<0.
\end{eqnarray*}

\end{document}